\documentclass[11pt,twocolumn]{article}
\usepackage{graphicx}
\usepackage{amsmath}
\usepackage{bm}

\begin{document}

\title{\textsf{\textbf{Poisson commutator-anticommutator brackets for ray tracing and longitudinal imaging via geometric algebra}}}
\author{Quirino M. Sugon Jr.* and Daniel J. McNamara
\smallskip\\
\small{Ateneo de Manila University, Department of Physics, Loyola Heights, Quezon City, Philippines 1108}\\
\small{*Also at Manila Observatory, Upper Atmosphere Division, Ateneo de Manila University Campus}\\
\small{e-mail: \texttt{qsugon$@$observatory.ph}}}
\date{\small{\today}}
\maketitle

\small
\section*{}\label{Abstract}
\textbf{Abstract.}  
We use the vector wedge product in geometric algebra to show that Poisson commutator brackets measure preservation of phase space areas.  We also use the vector dot product to define the Poisson anticommutator bracket that measures the preservation of phase space angles.  We apply these brackets to the paraxial meridional complex height-angle ray vectors that transform via a $2\times 2$ matrix, and we show that this transformation preserves areas but not angles in phase space.  The Poisson brackets here are expressed in terms of the coefficients of the $ABCD$ matrix.  We also apply these brackets to the distance-height ray vectors measured from the input and output side of the optical system.  We show that these vectors obey a partial Moebius transformation, and that this transformation preserves neither areas nor angles.  The Poisson brackets here are expressed in terms of the transverse and longitudinal magnifications.     

\section{Introduction}

\textbf{a.  Poisson Brackets.}  In two dimensional space, the Poisson bracket\cite{MarsdenRatiu_1994_IntroductiontoMechanicsandSymmetry_p4} is defined by the commutator or the Jacobian
\begin{equation}
\label{eq:Poisson bracket commutator F G wrt q p intro}
[F,G]_{q,p}=\frac{\partial F}{\partial q}\frac{\partial G}{\partial p}-\frac{\partial F}{\partial p}\frac{\partial G}{\partial q}=
\left |
\begin{matrix}
\ \displaystyle\frac{\partial F}{\partial q}&\displaystyle\frac{\partial F}{\partial p}\ \\
\\
\ \displaystyle\frac{\partial G}{\partial q}&\displaystyle\frac{\partial G}{\partial p}\ 
\end{matrix}
\right|,
\end{equation}
where we used the subscript notation of Hand and Finch\cite{HandFinch_1998_AnalyticalMechanics_p217}.
If this bracket is unity, then the mapping from the object parallelogram $dq\wedge dp$ to its image $dF\wedge dG$ is a symplectic (area-preserving) map.\cite{Gole_2008_ScholarpediaSymplecticMaps}  

So we ask: if the Poisson bracket commutator in Eq.~(\ref{eq:Poisson bracket commutator F G wrt q p intro}) is a measure of area preservation, can we also define a similar bracket for the preservation of the interior angles of the differential parallelograms $dq\wedge dp$ and $dF\wedge dG$?  Since this problem appears not to be taken before, we shall propose the following anticommutator bracket measure:
\begin{equation}
\label{eq:Poisson bracket anticommutator F G wrt q p intro}
\{F, G\}_{q,p}=\frac{\partial F}{\partial q}\frac{\partial F}{\partial p}+\frac{\partial G}{\partial p}\frac{\partial G}{\partial q},
\end{equation}     
which we shall show later to be based on the definition of the dot or inner vector product.  If $dq\perp dp$ and the bracket in Eq.~(\ref{eq:Poisson bracket anticommutator F G wrt q p intro}) is zero, then $dF\perp dG$. 

In geometric algebra, the wedge (outer) and dot (inner) products of two vectors $\mathbf a$ and $\mathbf b$ are related by their geometric (juxtaposition) product\cite{Hestenes_2003_ajpv71i2pp104-121_p107}:
\begin{equation}
\label{eq:ab is a dot b + a wedge b intro}
\mathbf a\mathbf b=\mathbf a\cdot\mathbf b+\mathbf a\wedge\mathbf b.
\end{equation}
The dot product is the same as in the standard vector analysis; the wedge product is the same as the Grassman outer product\cite{Doran_1993_jmathphysv4i8_pp3642-3669_sec2eq1-2}.  Doran and Lasenby used the wedge product of two vectors to express the Poisson commutator bracket commutator in Eq.~(\ref{eq:Poisson bracket commutator F G wrt q p intro}) as\cite{DoranLasenby_2003_GeometricAlgebraforPhysicists_p436}
\begin{equation}
\label{eq:F commutator G Poisson geometric algebra intro}
[\,F,G\,]=(\nabla F\wedge\nabla G)\cdot J,
\end{equation}
where $J$ is a bivector (two form or oriented area); however, they have not used the dot product to the define the bracket's anticommutator counterpart.  

\textbf{b. Paraxial Optics.}  In paraxial meridional ray optics, a light ray is described by the height angle vector $(x,n\alpha)$,
where $x$ is the height of the light particle from the optical axis, $n$ is the refractive index of the medium, and $\alpha$ is the inclination angle of the direction of propagation of light.  The input (unprimed) and output (primed) light rays are related by the $ABCD$ system matrix:\cite{NazarathyShamir_1982_josav72i3pp356-364_p358}
\begin{equation}
\label{eq:r is M r intro}
\begin{pmatrix}
x'\\
n'\alpha'
\end{pmatrix}
=
\begin{pmatrix}
A&B\\
C&D
\end{pmatrix}
\begin{pmatrix}
x\\
n\alpha
\end{pmatrix}.
\end{equation}
Though this linear transformation is known to preserve the area $dx\wedge d(n\alpha)$, we shall show that this transformation does generally preserve the angle between $dx$ and $d(n\alpha)$.

For imaging systems, the position of the object is given by $(S, x)$ and that of the image by $(S',x')$, where $S=s/n$ is the reduced distance of the object measured from the input side of the optical system, and $S'=s'/n'$ is that of the image measured from the output side.  These parameters are related by the partial Moebius relations\cite{SugonMcNamara_2008_arXiv0812.0664v1_p8}
\begin{eqnarray}
\label{eq:S' is M_11S + M_21 over M_12S - M_22 intro}
S'&=&\frac{M_{11}S+M_{21}}{M_{12}S-M_{22}},\\
\label{eq:x' is -x over M_12S - M_22}
x'&=&\frac{-x}{M_{12}S-M_{22}},
\end{eqnarray}
where $M_{11}$, $M_{12}$, $M_{21}$, and $M_{22}$ are constants.  We shall show that the transformation described by Eqs.~(\ref{eq:S' is M_11S + M_21 over M_12S - M_22 intro}) and (\ref{eq:x' is -x over M_12S - M_22}) is not symplectic, because the area $dS\wedge dx$ is not preserved.  We shall also show that this transformation does not preserve angles.

\textbf{c.  Outline.}  We shall divide the paper into five sections.  The first section is Introduction.  In the second section, we shall present geometric algebra, we we shall discuss products of vectors and of complex vectors.  In the third section, we shall summarize the matrix equations in paraxial meridional ray tracing in complex vector formalism we proposed in our previous paper\cite{SugonMcNamara_2008_arXiv0812.0664v1_p8}.  We shall discuss the linear height-angle transformation and the Moebius-like distance-height transformation.  In the fourth section, we shall compute the Poisson commutator and anticommutator brackets of these transformations.  We shall show that the height-angle transformation preserves areas but not angles in phase space, while the distance-height transformation neither preserve areas nor angles of longitudinal objects.

\section{Geometric Algebra}

The Clifford (geometric) algebra $\mathcal Cl_{3,0}$ is a group\cite{SugonMcNamara_2008_arXiv0809.0351v1_p2-3} algebra over the field of real numbers.  The generators of the group are the three vectors ${\bf e}_1$, ${\bf e}_2$, and ${\bf e}_3$ that satisfy the orthonormality relation\cite{BaylisHuschiltWei_1992_ajpv60i9pp788-797_p789}
\begin{equation}
\label{eq:e_je_k + e_ke_j is 2 delta_jk}
{\bf e}_j{\bf e}_k + {\bf e}_k{\bf e}_j = 2\delta_{jk},
\end{equation}
for $j,k=1,2,3$.  That is, 
\begin{eqnarray}
\label{eq:e_j^2 is 1}
\mathbf e_j^2 &=& 1,\\ 
\label{eq:e_je_k is -e_ke_j}
\mathbf e_j\mathbf e_k &=& -\mathbf e_k\mathbf e_j,  
\end{eqnarray}
for $j\neq k$.  Equation~(\ref{eq:e_j^2 is 1}) is called the normality axiom: a vector with a unit square has a unit norm or length.  Equation~(\ref{eq:e_je_k is -e_ke_j}) is the orthogonality axiom: perpendicular vectors anticommute.

\subsection{Vector Products}  

Let ${\bf a}$ and ${\bf b}$ be two vectors spanned by the three unit spatial vectors in $\mathcal Cl_{3,0}$:
\begin{eqnarray}
\label{eq:a is a_1e_1 + a_2e_2 + a_3e_3}
\mathbf a&=&a_1\mathbf e_1+a_2\mathbf e_2+a_3\mathbf e_3,\\
\label{eq:b is b_1e_1 + b_2e_2 + b_3e_3}
\mathbf a&=&a_1\mathbf e_1+a_2\mathbf e_2+a_3\mathbf e_3.
\end{eqnarray}
By the orthonormality axiom in Eq. (\ref{eq:e_je_k + e_ke_j is 2 delta_jk}), we can show that the geometric (juxtaposed) product of these two vectors is
\begin{equation}
\label{eq:ab is a dot b + a wedge b}
\mathbf a\mathbf b=\mathbf a\cdot\mathbf b+\mathbf a\wedge\mathbf b,
\end{equation}
where
\begin{eqnarray}
\label{eq:a dot b}
\mathbf a\cdot\mathbf b&=&a_1b_1+a_2b_2+a_3b_3,\\
\label{eq:a wedge b}
\mathbf a\wedge\mathbf b&=&\mathbf e_1\mathbf e_2(a_1b_2-a_2b_1)+\mathbf e_2\mathbf e_3(a_2b_3-a_3b_2)\nonumber\\
& &+\ \mathbf e_3\mathbf e_1(a_3b_1-a_1b_3)
\end{eqnarray}
are the inner (dot) and outer (wedge) products of the two vectors\cite{Jancewicz_1988_MultivectorsandCliffordAlgebrainElectrodynamics_p8}.

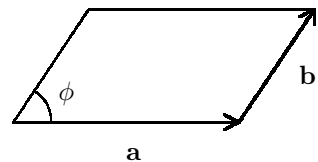
\begin{figure}[hb]
\begin{center}
\setlength{\unitlength}{1 mm}
\begin{picture}(40,30)(0,-10)
\thicklines
\qbezier(0,0)(30,0)(30,0)
\qbezier(28,1)(30,0)(30,0)
\qbezier(28,-1)(30,0)(30,0)
\put(15,-5){\small$\mathbf a$}
\qbezier(30,0)(40,15)(40,15)
\qbezier(38,14)(40,15)(40,15)
\qbezier(40,13)(40,15)(40,15)
\put(38,5){\small$\mathbf b$}
\thinlines
\qbezier(0,0)(10,15)(10,15)
\qbezier(10,15)(40,15)(40,15)
\qbezier(5.000,0.000)(5.000,2.676)(2.774,4.160)
\put(6,3){\small$\phi$}
\end{picture}
\end{center}
\begin{quote}
\vspace{-0.5cm}
\caption{\footnotesize An oriented area $\mathbf a\wedge\mathbf b$ defined by the vectors $\mathbf a$ and $\mathbf b$.  The angle between the two vectors is $\phi$.}
\label{fig:a wedge b with phi}
\vspace{-0.5cm}
\end{quote}
\end{figure}

To geometrically interpret the dot and wedge products of two vectors, let us define $\mathbf a$ as a vector along $\mathbf e_1$ and $\mathbf b$ as the vector along $\mathbf e_1$ rotated counterclockwise about the vector $\mathbf e_3$ by an angle $\phi$:
\begin{eqnarray}
\label{eq:a is ae_1}
\mathbf a&=&a\mathbf e_1,\\
\label{eq:b is e_1 b cos phi  + e_2 b sin phi}
\mathbf b&=&\mathbf e_1b\cos\phi+\mathbf e_2b\sin\phi.
\end{eqnarray}
The product of $\mathbf a$ and $\mathbf b$ is
\begin{equation}
\label{eq:ab is ab cos phi + e_1e_2 ab sin phi}
\mathbf a\mathbf b=ab\cos\phi+\mathbf e_1\mathbf e_2ab\sin\phi.
\end{equation}
Separating the scalar and bivector parts, we get
\begin{eqnarray}
\label{eq:a dot b is ab cos phi}
\mathbf a\cdot\mathbf b&=&ab\cos\phi,\\
\label{eq:a wedge b is ab sin phi e_1e_2}
\mathbf a\wedge\mathbf b&=&i\mathbf e_3ab\sin\phi=\mathbf e_1\mathbf e_2ab\sin\phi. 
\end{eqnarray}
Thus, the scalar $\mathbf a\cdot\mathbf b$ is the product the magnitude $a$ of vector $\mathbf a$ and the component of $b\cos\phi$ of the vector $\mathbf b$ along $\mathbf a$; the bivector $\mathbf a\wedge\mathbf b$ is proportional to the area of the parallelogram defined by vectors $\mathbf a$ and $\mathbf b$. (Fig. \ref{fig:a wedge b with phi})  
\medskip

The wedge product $\mathbf a\wedge\mathbf b$ in Eq.~(\ref{eq:a wedge b}) may be expressed in determinant form as
\begin{equation}
\label{eq:a wedge b as determinant}
\mathbf a\wedge\mathbf b=
\left|\ 
\begin{matrix}
a_1 & b_1\\
a_2 & b_2\\
\end{matrix}
\right|\mathbf e_1\mathbf e_2+
\left|\ 
\begin{matrix}
a_2 & b_2\\
a_3 & b_3\\
\end{matrix}
\right|\mathbf e_2\mathbf e_3+
\left|\ 
\begin{matrix}
a_3 & b_3\\
a_1 & b_1\\
\end{matrix}
\right|\mathbf e_3\mathbf e_1.
\end{equation}
Using the properties of determinants, we can show that 
\begin{eqnarray}
\label{eq:a wedge b is -b wedge a}
\mathbf a\wedge\mathbf b&=&-\mathbf b\wedge\mathbf a,\\
\label{eq:a wedge b is 0}
\mathbf a\wedge\mathbf a&=&0.
\end{eqnarray}
The first equation states that the orientation of the directed area defined by two vectors flips if the vector factors are interchanged.  The second equation states that no area can be defined by a vector and itself.

The dot product is a grade-lowering operation; the wedge product is a grade-raising operation.  That is,
\begin{eqnarray}
\label{eq:a dot b is ab_1-1 component}
\mathbf a\cdot\mathbf b =\langle\mathbf a\mathbf b\rangle_{|1-1|}=\langle\mathbf a\mathbf b\rangle_0,\\
\label{eq:a wedge b is ab_1+1 component}
\mathbf a\wedge\mathbf b =\langle\mathbf a\mathbf b\rangle_{|1+1|}=\langle\mathbf a\mathbf b\rangle_2,
\end{eqnarray}
where the angle brackets $\langle\mathbf a\mathbf b \rangle_{|g|}$ denotes the extraction the $g-$vector part of $\mathbf a\mathbf b$.  In other words, the grade of $\mathbf a\cdot\mathbf b$ is the sum of the grades of $\mathbf a$ and $\mathbf b$; the grade of $\mathbf a\wedge\mathbf b$ is the difference of the grades of $\mathbf a$ and $\mathbf b$.

\subsection{Hodge Map and Spatial Inversion}  

Let us denote the product of the three orthonormal vectors $\mathbf e_1$, $\mathbf e_2$, and $\mathbf e_3$ by $i$:\cite{BaylisHuschiltWei_1992_ajpv60i9pp788-797_p790}
\begin{equation}
\label{eq:i = e_1e_2e_3}
i=\mathbf e_1\mathbf e_2\mathbf e_3.
\end{equation}
The geometric interpretation of $i$ is an oriented volume.

We chose the notation $i$ because the trivector behaves like a true unit imaginary scalar.  That is, its square is $-1$,
\begin{eqnarray}
\label{eq:i^2 = -1}
i^2&=&\mathbf e_1\mathbf e_2\mathbf e_3\mathbf e_1\mathbf e_2\mathbf e_3=\mathbf e_1^2\mathbf e_2\mathbf e_3\mathbf e_2\mathbf e_3\nonumber\\
&=&-\mathbf e_1^2\mathbf e_2^2\mathbf e_3^2=-1,
\end{eqnarray}
by the orthonormality axiom in Eq.~(\ref{eq:e_je_k + e_ke_j is 2 delta_jk}), and that it commutes with vectors,
\begin{eqnarray}
\label{eq:ie_1 is e_2e_3 is e_1i}
i\mathbf e_1 &=& \mathbf e_2\mathbf e_3 = \mathbf e_1i,\\
\label{eq:ie_2 is e_3e_1 is e_2i}
i\mathbf e_2 &=& \mathbf e_3\mathbf e_1 = \mathbf e_2i,\\
\label{eq:ie_3 is e_1e_2 is e_3i}
i\mathbf e_3 &=& \mathbf e_1\mathbf e_2 = \mathbf e_3i.
\end{eqnarray}
Notice that the trivector $i$ transforms a vector into a bivector (and a bivector into a vector).  This transformation is called a Hodge map or a duality transformation.

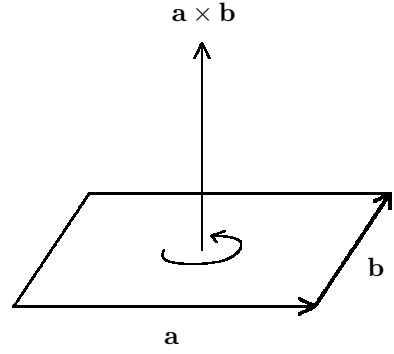
\begin{figure}[ht]
\label{fig:a x b is i a wedge b}
\begin{center}
\setlength{\unitlength}{1 mm}
\begin{picture}(50,55)(0,-10)
\thicklines
\qbezier(0,0)(40,0)(40,0)
\qbezier(38,1)(40,0)(40,0)
\qbezier(38,-1)(40,0)(40,0)
\put(20,-5){\small$\mathbf a$}
\thicklines
\qbezier(40,0)(50,15)(50,15)
\qbezier(50,13)(50,15)(50,15)
\qbezier(48,14)(50,15)(50,15)
\put(47,4){\small$\mathbf b$}
\thinlines
\qbezier(0,0)(10,15)(10,15)
\qbezier(10,15)(50,15)(50,15)
\qbezier(25,7.5)(25,35)(25,35)
\qbezier(24,33)(25,35)(25,35)
\qbezier(26,33)(25,35)(25,35)
\put(21,38){\small$\mathbf a\times\mathbf b$}
\qbezier(20,7.5)(18.75,5.625)(23.75,5.625)
\qbezier(23.75,5.625)(28.75,5.625)(30,7.5)
\qbezier(30,7.5)(31.25,9.375)(26.25,9.375)
\qbezier(27,8.5)(26.25,9.375)(26.25,9.375)
\qbezier(28,10)(26.25,9.375)(26.25,9.375)
\end{picture}
\end{center}
\begin{quote}
\vspace{-0.5cm}
\caption{\footnotesize The oriented plane $\mathbf a\wedge\mathbf b$ and its normal vector $\mathbf a\times\mathbf b$.}
\label{fig:a wedge b and a x b}
\vspace{-0.5cm}
\end{quote}
\end{figure}

Using the relations in Eq.~(\ref{eq:ie_1 is e_2e_3 is e_1i}) to (\ref{eq:ie_3 is e_1e_2 is e_3i}), we can show that the wedge product $\mathbf a\wedge\mathbf b$ is related to the cross product $\mathbf a\times\mathbf b$ by a Hodge map:
\begin{equation}
\label{eq:a wedge b is i a x b}
\mathbf a\wedge\mathbf b=i(\mathbf a\times\mathbf b),
\end{equation}
where
\begin{eqnarray}
\label{eq:a x b expand}
\mathbf a\times\mathbf b &=& (a_2b_3-a_3b_2)\mathbf e_1+(a_3b_1-a_1b_3)\mathbf e_2\nonumber\\
&\,&+(a_1b_2-a_2b_1)\mathbf e_3.
\end{eqnarray}
Geometrically, the bivector $\mathbf a\wedge\mathbf b$ is the oriented plane with $\mathbf a\times\mathbf b$ as its associated normal vector.  (Fig. \ref{fig:a wedge b and a x b})

Substituting the Hodge map relation in Eq.~(\ref{eq:a wedge b is i a x b}) back to the juxtaposed product expansion in Eq.~(\ref{eq:ab is a dot b + a wedge b}), we get\cite{Hestenes_2003_ajpv71i2pp104-121_p110}
\begin{equation}
\label{eq:ab is a dot b + i axb}
\mathbf a\mathbf b=\mathbf a\cdot\mathbf b + i(\mathbf a\times\mathbf b).
\end{equation}
This equation is similar to the Pauli identity in Quantum Mechanics, except for the absence of the Pauli $\hat\sigma-$matrices.

In general, every element $\hat A$ in $\mathcal Cl_{3,0}$ may be expressed as a sum of a scalar, a vector, a bivector (imaginary vector), and a trivector (imaginary scalar):\cite{Hestenes_2003_ajpv71i2pp104-121_p110}
\begin{equation}
\label{eq:A}
\hat A=A_0+\mathbf A_1+i\mathbf A_2+iA_3.
\end{equation}
We call such a sum a cliffor.  The spatial inverse of the cliffor $\hat A$ is\cite{SugonMcNamara_2008_arXiv:0807.1382v1_p2} 
\begin{equation}
\label{eq:A dagger is A_0 - A_1 + iA_2 - iA_3}
\hat A^\dagger =A_0-\mathbf A_1+i\mathbf A_2-iA_3.
\end{equation}
That is, the spatial inversion operator $(^\dagger)$ flips the signs of vectors and trivectors, but leaves scalars and bivectors unchanged.  The other term for spatial inversion is automorphic grade involution\cite{Baylis_1996_CliffordGeometricAlgebras_p4}.

Other properties of the spatial inversion operator are as follows:
\begin{eqnarray}
\label{eq:A + B dagger distribute}
(\hat A+\hat B)^\dagger&=&\hat A^\dagger+\hat B^\dagger,\\
\label{eq:AB dagger distribute}
(\hat A\hat B)^\dagger&=&\hat A^\dagger\hat B^\dagger,\\
\label{eq:A dagger dagger is A}
(\hat A^\dagger)^\dagger&=&\hat A.
\end{eqnarray}
These properties can be derived from the definition of the spatial inverse of $\hat A$ in terms of the unit vector $\mathbf e_0=\mathbf e_4$ in $\mathcal Cl_{4,0}$ that is orthogonal to $\mathbf e_1$, $\mathbf e_2$, and $\mathbf e_3$:\cite{SugonMcNamara_2008_arXiv:0807.1382v1_p2}
\begin{equation}
\label{eq:Ae_0 is e_0A dagger}
\hat A\mathbf e_0=\mathbf e_0\hat A^\dagger.
\end{equation}
Notice that spatial inversion does not flip the order of the products, unlike in the case of the reversion \cite{Vold_1993_ajp61i6pp491-504_p499} (though they share the same dagger notation).  

\subsection{Complex Vector Products}

Let $\hat r$ and $\hat r'$ be complex vectors:
\begin{eqnarray}
\label{eq:r is a + ib}
\hat r&=&\mathbf a+i\mathbf b,\\
\label{eq:r' is c + id}
\hat r'&=&\mathbf c+i\mathbf d.
\end{eqnarray}
The product of $\hat r$ and $\hat r'$ is
\begin{equation}
\label{eq:rr' is a + ib and c + id}
\hat r\hat r'=(\mathbf a+i\mathbf b)(\mathbf c+i\mathbf d).
\end{equation}
Separating the scalar, vector, bivector, and trivector parts of Eq.~(\ref{eq:rr' is a + ib and c + id}), we get
\begin{eqnarray}
\label{eq:rr'_0}
\langle\hat r\hat r'\rangle_0&=&\mathbf a\cdot\mathbf c-\mathbf b\cdot\mathbf d,\\
\label{eq:rr'_1}
\langle\hat r\hat r'\rangle_1&=&-\mathbf a\times\mathbf d-\mathbf b\times\mathbf c,\\
\label{eq:rr'_2}
\langle\hat r\hat r'\rangle_2&=&i(\mathbf a\times\mathbf c-\mathbf b\times\mathbf d),\\
\label{eq:rr'_3}
\langle\hat r\hat r'\rangle_3&=&i(\mathbf a\cdot\mathbf d+\mathbf b\cdot\mathbf c).
\end{eqnarray}
Take note of Eq.~(\ref{eq:rr'_1}).

On the other hand, the product of $\hat r$ and $\hat r'^\dagger$ is
\begin{equation}
\label{eq:rr' dagger is a + ib and -c + id}
\hat r\hat r'^\dagger=(\mathbf a+i\mathbf b)(-\mathbf c+i\mathbf d).
\end{equation}
Separating the scalar, vector, bivector, and trivector parts of Eq.~(\ref{eq:rr' dagger is a + ib and -c + id}), we get
\begin{eqnarray}
\label{eq:rr' dagger_0}
\langle\hat r\hat r'^\dagger\rangle_0&=&-\mathbf a\cdot\mathbf c-\mathbf b\cdot\mathbf d,\\
\label{eq:rr' dagger_1}
\langle\hat r\hat r'^\dagger\rangle_1&=&-\mathbf a\times\mathbf d+\mathbf b\times\mathbf c,\\
\label{eq:rr' dagger_2}
\langle\hat r\hat r'^\dagger\rangle_2&=&i(-\mathbf a\times\mathbf c-\mathbf b\times\mathbf d),\\
\label{eq:rr' dagger_3}
\langle\hat r\hat r'^\dagger\rangle_3&=&i(\mathbf a\cdot\mathbf d-\mathbf b\cdot\mathbf c).
\end{eqnarray}
Take note of Eq.~(\ref{eq:rr' dagger_0}).

To clarify the geometric interpretations of Eqs.~(\ref{eq:rr'_1}) and (\ref{eq:rr' dagger_0}), let us go to the complex vector phase space and replace the imaginary parts by vectors. (Though this procedure is strictly not allowed, it is still pedagogically instructive.)  That is, we write
\begin{eqnarray}
\label{eq:r is a + b}
\mathbf r=\mathbf a+\mathbf b,\\
\label{eq:r' is c + d}
\mathbf r'=\mathbf c+\mathbf d.
\end{eqnarray}
The product of the vectors $\mathbf r$ and $\mathbf r'$ is
\begin{equation}
\label{eq:rr' is a + b and c + d}
\mathbf r\mathbf r'=(\mathbf a+\mathbf b)(\mathbf c+\mathbf d).
\end{equation}
Separating the scalar and bivector parts of Eq.~(\ref{eq:rr' is a + b and c + d}) yields
\begin{eqnarray}
\label{eq:rr'_0 phase}
\mathbf r\cdot\mathbf r'&=&\mathbf a\cdot\mathbf c+\mathbf a\cdot\mathbf d+\mathbf b\cdot\mathbf c+\mathbf b\cdot\mathbf d,\\
\label{eq:rr'_1 phase}
i(\mathbf r\times\mathbf r')&=&i(\mathbf a\times\mathbf c+\mathbf a\times\mathbf d+\mathbf b\times\mathbf c+\mathbf b\times\mathbf d).
\end{eqnarray}
Note that the scalar part is proportional to the cosine of the angle between $\mathbf r$ and $\mathbf r'$, while the magnitude of the bivector part is the area defined by $\mathbf r$ and $\mathbf r'$.

\begin{figure}[hb]
\begin{center}
\setlength{\unitlength}{1 mm}
\begin{picture}(50,55)(0,-10)
\multiput(0,0)(2,0){15}{\line(1,0){1}}
\qbezier(28,0)(30,0)(30,0)
\qbezier(28,-1)(30,0)(30,0)
\qbezier(28,1)(30,0)(30,0)
\put(15,-5){\small$\mathbf a$}
\multiput(30,0)(0,2){5}{\line(0,1){1}}
\qbezier(30,8)(30,10)(30,10)
\qbezier(31,8)(30,10)(30,10)
\qbezier(29,8)(30,10)(30,10)
\put(32,2){\small$\mathbf b$}
\multiput(30,10)(2,0){7}{\line(1,0){1}}
\qbezier(43,10)(45,10)(45,10)
\qbezier(43,11)(45,10)(45,10)
\qbezier(43,9)(45,10)(45,10)
\put(40,6){\small$\mathbf c$}
\multiput(45,10)(0,2){15}{\line(0,1){1}}
\qbezier(45,38)(45,40)(45,40)
\qbezier(46,38)(45,40)(45,40)
\qbezier(44,38)(45,40)(45,40)
\put(47,23){\small$\mathbf d$}
\thicklines
\qbezier(0,0)(30,10)(30,10)
\qbezier(27.5,10.5)(30,10)(30,10)
\qbezier(30,10)(45,40)(45,40)
\qbezier(43,39)(45,40)(45,40)
\thinlines
\qbezier(0,0)(15,30)(15,30)
\qbezier(15,30)(45,40)(45,40)
\end{picture}
\end{center}
\begin{quote}
\vspace{-0.5cm}
\caption{\footnotesize The oriented area $(\mathbf a+\mathbf b)\wedge(\mathbf c+\mathbf d)$.  This only serves as a way of visualizing the phase plane defined by the complex vectors $\hat r=\mathbf a+i\mathbf b$ and $\hat r=\mathbf c+i\mathbf d$.}
\label{fig:a + b wedge c + d}
\vspace{-0.5cm}
\end{quote}
\end{figure}
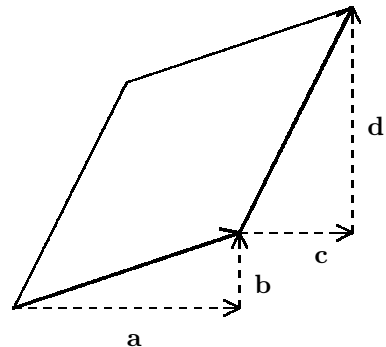

If we assume that $\mathbf a\parallel\mathbf c$, $\mathbf b\parallel\mathbf d$, and $\mathbf a\perp\mathbf b$, then
\begin{eqnarray}
\label{eq:a dot d is b dot c is a times c is b times d is 0}
\mathbf a\cdot\mathbf d=\mathbf b\cdot\mathbf c=\mathbf a\times\mathbf c=\mathbf b\times\mathbf d=0.
\end{eqnarray}
Substituting these results back to Eqs.~(\ref{eq:rr'_0 phase}) and (\ref{eq:rr'_1 phase}), we obtain
\begin{eqnarray}
\label{eq:rr'_0 phase is a dot c + b dot d}
\mathbf r\cdot\mathbf r'&=&\mathbf a\cdot\mathbf c+\mathbf b\cdot\mathbf d,\\
\label{eq:rr'_2 phase is i a times d + b times c}
i(\mathbf r\times\mathbf r')&=&i(\mathbf a\times\mathbf d+\mathbf b\times\mathbf c).
\end{eqnarray}
Notice that $\mathbf r\cdot\mathbf r'$ is the negative of that $\langle\hat r\hat r'^\dagger\rangle_0$ in Eq.~(\ref{eq:rr' dagger_0}), and that the area $|\mathbf r\times\mathbf r'|$ is the same as the magnitude of $\langle\hat r\hat r'\rangle_1$ in Eq.~(\ref{eq:rr'_1}).  Later, we shall use these results later when we define the Poisson commutator and anticommutator brackets in complex vector spaces.  (Fig.~\ref{fig:a + b wedge c + d})

\section{Matrix Optics}

\subsection{Height-Angle Relations}

\begin{figure}[hb]
\begin{center}
\setlength{\unitlength}{1 mm}
\begin{picture}(60,60)(0,-10)
\put(0,0){\line(0,1){15}}
\put(2,6){\small $x_1$}
\put(30,0){\line(0,1){35}}
\put(32,15){\small $x_2$}
\multiput(55,0)(0,5){9}{\line(0,1){2.5}}
\put(57,20){\small $x_3$}
\put(0,0){\line(1,0){30}}
\put(15,-5){\small $s_1$}
\multiput(32.5,0)(5,0){5}{\line(1,0){2.5}}
\put(42,-5){\small $s_2$}
\thicklines
\qbezier(0,15)(30,35)(30,35)
\thinlines
\put(0,15){\line(1,0){10}}
\qbezier(7.000,15.000)(7.000,17.119)(5.824,18.883)
\put(9,17){\small $\alpha_1$}
\thicklines
\qbezier(30,35)(55,45)(55,45)
\thinlines
\qbezier(37.000,35.000)(37.000,36.348)(36.499,37.600)
\put(30,35){\line(1,0){10}}
\put(40,36){\small $\alpha_2$}
\end{picture}
\end{center}
\begin{quote}
\vspace{-0.5cm}
\caption{\footnotesize A paraxial ray with height $x_1$ and inclination angle $\alpha_1$ moves to the right by a distance $s_1$ until it hits an refracting surface.  The height of the ray becomes $x_2$ and its inclination angle changes to $\alpha_2$.}
\label{fig:propagation refraction propagation}
\vspace{-0.5cm}
\end{quote}
\end{figure}
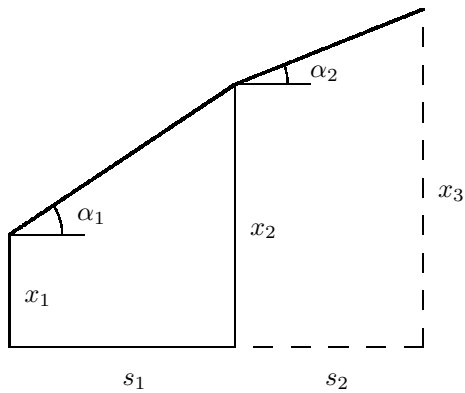

In matrix optics, a height-angle ray vector $\hat r$ may be  described by a complex column vector, which we shall interpret in terms of the orthonormal vectors $\mathbf e_1$ pointing upward and $\mathbf e_2$ pointing out of the paper:\cite{SugonMcNamara_2008_arXiv:0812.0664v1_p3}
\begin{equation}
\label{eq:r is r e_1 + inalpha e_2}
\hat r=
\begin{pmatrix}
x\\
in\alpha
\end{pmatrix}
=x\mathbf e_1+in\alpha\mathbf e_2,
\end{equation}
where $x$ is the height of the light particle with respect to the optical axis $\mathbf e_3$ pointing to the right, $n$ is the refractive index of the medium containing the light particle, and $\alpha$ is the counterclockwise paraxial angle of inclination of the direction of propagation of the light particle (Fig. \ref{fig:propagation refraction propagation}).  Note that the imaginary number $i=\mathbf e_1\mathbf e_2\mathbf e_3$ is a trivector, and $i\mathbf e_2=\mathbf e_3\mathbf e_1$ is the $zx-$plane (bivector) containing the angle $n\alpha$. 

Let $\textsf M$ be a complex system matrix,
\begin{equation}
\label{eq:M matrix}
\textsf M=
\begin{pmatrix}
A&-iC\\
-iB&D
\end{pmatrix},
\end{equation}
and let its determinant be unity,
\begin{equation}
\label{eq:det M is AD + BC is 1}
|\textsf M|=AD + BC=1.
\end{equation}
Notice that because of the presence of the imaginary numbers, the determinant is not a difference, as given in the literature, but a sum.

When a light particle, characterized by its height-angle ray vector $\hat r$, enters the (left) input side  of an optical system described by the right-acting matrix $\textsf M$, 
the height-angle ray vector of the light particle changes into $\hat r'$ at the (right) output side.  In other words,
\begin{equation}
\label{eq:r' is M^T r is r M}
\hat r'=\textsf M^T\cdot\hat r=\hat r\cdot\textsf M.
\end{equation}
That is,\cite{SugonMcNamara_2008_arXiv:0812.0664v1_p4}
\begin{eqnarray}
\label{eq:r' is M^T r is r M matrix}
\begin{pmatrix}
x'\\
in\alpha'
\end{pmatrix}
&=&
\begin{pmatrix}
A&-iB\\
-iC&D
\end{pmatrix}
\begin{pmatrix}
x\\
in\alpha
\end{pmatrix}
\nonumber\\
&=&
\begin{pmatrix}
x\\
in\alpha
\end{pmatrix}
\begin{pmatrix}
A&-iC\\
-iB&D
\end{pmatrix}
.
\end{eqnarray}
Separating the $\mathbf e_1$ and $i\mathbf e_2$ components of Eq.~(\ref{eq:r' is M^T r is r M matrix}), we get
\begin{eqnarray}
\label{eq:x' is Ax + Bnalpha}
x^\prime&=&Ax+Bn\alpha\\
\label{eq:n'alpha' is -Cx + Dnalpha}
n^\prime\alpha^\prime&=&-Cx+Dn\alpha.
\end{eqnarray}
Except for the sign of $C$, these equations are similar to those in the literature.

\subsection{Distance-Height Relations}

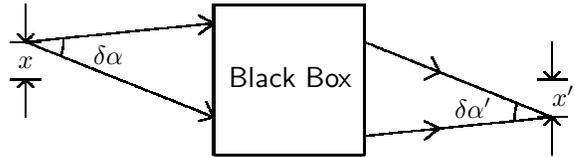
\begin{figure}[hb]
\begin{center}
\setlength{\unitlength}{1 mm}
\begin{picture}(70,35)(-25,-5)
\thicklines
\multiput(0,0)(0,20){2}{\line(1,0){20}}
\multiput(0,0)(20,0){2}{\line(0,1){20}}
\put(2,9){\textsf{Black Box}}
\thinlines
\qbezier(-25,15)(0,17.5)(0,17.5)
\qbezier(-2,18.5)(0,17.5)(0,17.5)
\qbezier(-2,16)(0,17.5)(0,17.5)
\qbezier(-25,15)(0,5)(0,5)
\qbezier(-2,4.5)(0,5)(0,5)
\qbezier(-1,7)(0,5)(0,5)
\qbezier(-20.025,15.498)(-19.903,14.280)(-20.358,13.143)
\put(-16,12){\small{$\delta\alpha$}}
\multiput(-25,5)(0,10){2}{\line(0,1){5}}
\qbezier(-26,16)(-25,15)(-25,15)
\qbezier(-24,16)(-25,15)(-25,15)
\multiput(-27,10)(0,5){2}{\line(1,0){4}}
\qbezier(-26,9)(-25,10)(-25,10)
\qbezier(-24,9)(-25,10)(-25,10)
\put(-26,11.5){\small{$x$}}
\thinlines
\qbezier(20,2.5)(45,5)(45,5)
\qbezier(28,4.5)(30,3.5)(30,3.5)
\qbezier(28,2)(30,3.5)(30,3.5)
\qbezier(20,15)(45,5)(45,5)
\qbezier(29,13)(30,11)(30,11)
\qbezier(28,10)(30,11)(30,11)
\qbezier(40.025,4.502)(39.903,5.720)(40.358,6.857)
\put(32,5){\small{$\delta\alpha'$}}
\multiput(45,0)(0,10){2}{\line(0,1){5}}
\multiput(43,5)(0,5){2}{\line(1,0){4}}
\qbezier(44,4)(45,5)(45,5)
\qbezier(46,4)(45,5)(45,5)
\qbezier(44,11)(45,10)(45,10)
\qbezier(46,11)(45,10)(45,10)
\put(45,6.5){\small{$x'$}}

\end{picture}
\end{center}
\begin{quote}
\vspace{-0.5cm}
\caption{\footnotesize In an imaging system, rays leaving a point source converge to a point.  The distance of the object to the left input side of th black box is $S$, while that of the image from the right output side is $S'$.}
\label{fig:imaging system}
\vspace{-0.5cm}
\end{quote}
\end{figure}

Let $\mathbf r$ be the position of an object point measured from the input side of the black box and let $\mathbf r'$ be the image point measured from the output side (Fig. \ref{fig:imaging system}):
\begin{eqnarray}
\label{eq:r is Se_3 + xe_1}
\mathbf r&=&-S\mathbf e_3+x\mathbf e_1,\\
\label{eq:r' is S'e_3 + x'e_1}
\mathbf r'&=&S'\mathbf e_3+x'\mathbf e_1,
\end{eqnarray}
where 
\begin{eqnarray}
\label{eq:S is s/n}
S&=&s/n,\\
\label{eq:S' is s'/n}
S'&=&s'/n'
\end{eqnarray}
are the reduced object and image distances, while $x$ and $x'$ are the object and image heights. At these two points the complex height-angle ray vectors are given by $\hat r$ and $\hat r'$, respectively. 

Let us decompose the system matrix $\hat M$ as the product of a propagation matrix $\textsf S$, a box matrix $\textsf M_{\textrm{box}}$, and $\textsf M_S'$:\cite{SugonMcNamara_2008_arXiv0812.0664v1_p6}
\begin{equation}
\label{eq:M is M_S M_box M_S'}
\textsf M=\textsf M_S\textsf M_{box}\textsf M_{S'},
\end{equation}
where
\begin{eqnarray}
\label{eq:M_S}
\textsf M_S&=&
\begin{pmatrix}
1&0\\
-iS&1
\end{pmatrix}
\\
\label{eq:M_box}
\textsf M_{\textrm{box}}&=&
\begin{pmatrix}
M_{11}&-iM_{12}\\
-iM_{21}&M_{22}
\end{pmatrix}
\\
\label{eq:M_S'}
\textsf M_{S'}&=&
\begin{pmatrix}
1&0\\
-iS'&1
\end{pmatrix}
\end{eqnarray}
Notice that since $\textsf M$, $\textsf M_S$, and $\textsf M_{S'}$ have a unit determinant, then $\textsf M_{\textrm{box}}$ must also have a unit determinant:
\begin{equation}
\label{eq:det M_box is 1}
|\textsf M_{\textrm{box}}|=M_{11}M_{22}+M_{12}M_{21}.
\end{equation}

The transpose of Eq.~(\ref{eq:M is M_S M_box M_S'}) is
\begin{equation}
\label{eq:M^T is M_S^T M_box^T M_S'^T}
\textsf M^T=\textsf M_{S'}^T\textsf M_{\textrm{box}}^T\textsf M_S^T.
\end{equation}
Taking the transpose of the matrix $\textsf M$ in Eq.~(\ref{eq:M matrix}) and substituting the result in Eq.~(\ref{eq:M^T is M_S^T M_box^T M_S'^T}), we arrive at
\begin{eqnarray}
\label{eq:A element}
A&=&M_{11}-M_{12}S',\\
\label{eq:B element}
B&=&M_{21}+M_{22}S' +M_{11}S-M_{12}SS',\\
\label{eq:C element}
C&=&M_{12},\\
\label{eq:D element}
D&=&M_{22}-M_{12}S,
\end{eqnarray}
after carrying out the matrix products and separating the matrix coefficients.

Now, if $\mathbf r$ and $\mathbf r'$ in Eqs.~(\ref{eq:r is Se_3 + xe_1}) and (\ref{eq:r' is S'e_3 + x'e_1}) are positions of the object and image points, then the object and image heights $x$ and $x'$ are constants, so that the output ray angle $\alpha'$ depends only on the input ray angle $\alpha$:
\begin{equation}
\label{eq:alpha' is alpha' of alpha}
\alpha'=\alpha'(\alpha).
\end{equation}
Thus, taking the partial derivative of Eqs.~(\ref{eq:x' is Ax + Bnalpha}) and (\ref{eq:n'alpha' is -Cx + Dnalpha}) with respect to $n\alpha$,we get
\begin{eqnarray}
\label{eq:partial x' by partial nalpha is B is 0}
0&=&B,\\
\label{eq:partial n'alpha' by partial nalpha is D}
\frac{\partial(n'\alpha')}{\partial(n\alpha)}&=&D.
\end{eqnarray}
Our interest in this paper is only with the first equation.

Substituting Eq.~(\ref{eq:B element}) into Eqs.~(\ref{eq:x' is Ax + Bnalpha}) and (\ref{eq:partial x' by partial nalpha is B is 0}) and solving for the reduced distance $S'$ and the image height $x'$, we obtain\cite{SugonMcNamara_2008_arXiv0812.0664v1_p8}
\begin{eqnarray}
\label{eq:r' e_3 part transform}
S'&=&\frac{M_{11}S+M_{21}}{M_{12}S-M_{22}}=-m_x(M_{11}S+M_{21}),\\
\label{eq:r' e_1 part transform}
x'&=&-\frac{x}{M_{12}S-M_{22}}=m_xx,
\end{eqnarray}
where
\begin{equation}
\label{eq:m_x is x' over x}
m_x=\frac{x'}{x}=\frac{-1}{M_{12}S-M_{22}}=\frac{1}{D}.
\end{equation}
is the transverse magnification of the system.  Notice that the reduced output distance $S'$ is a Moebius transform of the reduced input distance $S$.

\section{Image Manifold}

\subsection{Height-Angle Phase Space}

\begin{figure}[hb]
\begin{center}
\setlength{\unitlength}{1 mm}
\begin{picture}(60,60)(-10,-5)
\put(0,0){\line(1,0){45}}
\qbezier(43,-1)(45,0)(45,0)
\qbezier(43,1)(45,0)(45,0)
\put(47,-1){\small$\mathbf e_1$}
\put(0,0){\line(0,1){45}}
\qbezier(-1,43)(0,45)(0,45)
\qbezier(1,43)(0,45)(0,45)
\put(-1,46){\small $i\mathbf e_2$}
\multiput(15,0)(0,2){8}{\line(0,1){1}}
\put(15,-5){\small$x$}
\multiput(0,15)(2,0){8}{\line(1,0){1}}
\put(-7,14){\small$n\alpha$}
\thicklines
\multiput(15,15)(0,25){2}{\line(1,0){25}}
\qbezier(39,14)(40,15)(40,15)
\qbezier(39,16)(40,15)(40,15)
\put(27,11){\small$dx$}
\multiput(15,15)(25,0){2}{\line(0,1){25}}
\qbezier(39,39)(40,40)(40,40)
\qbezier(41,39)(40,40)(40,40)
\put(42,25){\small$d(n\alpha)$}
\put(12,11){\small$\hat r_1$}
\put(41,11){\small$\hat r_2$}
\put(12,41){\small$\hat r_4$}
\put(41,41){\small$\hat r_3$}
\end{picture}
\end{center}
\begin{quote}
\vspace{-0.5cm}
\caption{\footnotesize An oriented area in phase space defined by $dx\,\mathbf e_1$ and $d(n\alpha)i\mathbf e_2$.}
\label{fig:dx dnalpha e_1ie_2}
\vspace{-0.5cm}
\end{quote}
\end{figure}
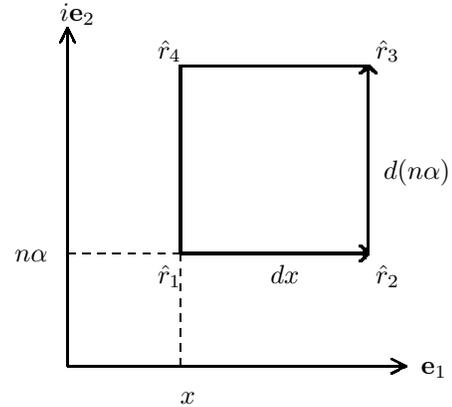

\textbf{a.  Object Phase Rectangle.}  In the height-angle phase space, let us construct a rectangle of height $dx$ and width $d(n\alpha)$:
\begin{eqnarray}
\label{eq:r_1 height angle}
\hat r_1&=&x\mathbf e_1+n\alpha i\mathbf e_2,\\
\label{eq:r_2 height angle}
\hat r_2&=&(x+dx)\mathbf e_1+n\alpha i\mathbf e_2,\\
\label{eq:r_3 height angle}
\hat r_3&=&(x+dx)\mathbf e_1+(n\alpha +d(n\alpha))i\mathbf e_2,\\
\label{eq:r_4 height angle}
\hat r_4&=&x\mathbf e_1+(n\alpha +d(n\alpha))i\mathbf e_2,
\end{eqnarray}
where $\hat r_1$, $\hat r_2$, $\hat r_3$, and $\hat r_4$ are consecutive vertices of the rectangle for a counterclockwise path.  (Fig. \ref{fig:dx dnalpha e_1ie_2})

The two consecutive sides of the rectangle starting from $\hat r_1$ are
\begin{eqnarray}
\label{eq:r_12 is dx e_1}
\hat r_{12}&=&\hat r_2-\hat r_1=dx\,\mathbf e_1,\\
\label{eq:r_23 is dnalpha ie_2}
\hat r_{23}&=&\hat r_3-\hat r_2=d(n\alpha)\,i\mathbf e_2.
\end{eqnarray}
The product of these two oriented sides is
\begin{equation}
\label{eq:r_12 r_23}
\hat r_{12}\hat r_{23}=-dx\,d(n\alpha)\mathbf e_3,
\end{equation}
because $\mathbf e_1i\mathbf e_2=-\mathbf e_3$.  Notice that Eq.~(\ref{eq:r_12 r_23}) is a pure vector, whose magnitude is the area of the phase space rectangle, as given by Eqs.~(\ref{eq:rr'_2 phase is i a times d + b times c}) and (\ref{eq:rr'_1}).  

On the other hand, the product of $\hat r_{12}$ and $\hat r_{23}^\dagger$ is
\begin{equation}
\label{eq:r_12 r_23 dagger}
\hat r_{12}\hat r_{23}^\dagger=dx\,d(n\alpha)\mathbf e_3,
\end{equation}
which is also a pure vector.  Notice that the vanishing of the scalar component means that the sides of the phase rectangle defined by $\hat r_{12}$ and $\hat r_{23}$ are indeed perpendicular, as given by Eqs.~(\ref{eq:rr' dagger_0}) and (\ref{eq:rr'_0 phase is a dot c + b dot d}).

\begin{figure}[hb]
\begin{center}
\setlength{\unitlength}{1 mm}
\begin{picture}(75,85)(-15,-25)
\put(-10,-15){\line(1,0){65}}
\qbezier(53,-14)(55,-15)(55,-15)
\qbezier(53,-16)(55,-15)(55,-15)
\put(57,-16){\small$\mathbf e_1$}
\put(-10,-15){\line(0,1){60}}
\qbezier(-11,43)(-10,45)(-10,45)
\qbezier(-9,43)(-10,45)(-10,45)
\put(-11,47){\small$i\mathbf e_2$}

\thicklines
\qbezier(0,0)(30,10)(30,10)
\qbezier(28.5,10.5)(30,10)(30,10)
\qbezier(29,8.5)(30,10)(30,10)

\qbezier(30,10)(45,40)(45,40)
\qbezier(45.5,38)(45,40)(45,40)
\qbezier(0,0)(15,30)(15,30)
\qbezier(15,30)(45,40)(45,40)
\put(10,-8){\footnotesize$\displaystyle\frac{\partial x'}{\partial x}\,dx$}
\put(31,-8){\footnotesize$\displaystyle\frac{\partial x'}{\partial(n\alpha)}\,d(n\alpha)$}
\put(48,4){\footnotesize$\displaystyle\frac{\partial (n'\alpha')}{\partial x}\,dx$}
\put(48,20){\footnotesize$\displaystyle\frac{\partial{n'\alpha'}}{\partial(n\alpha)}\,d(n\alpha)$}
\thinlines
\multiput(0,0)(2,0){24}{\line(1,0){1}}
\multiput(45,-2)(0,2){20}{\line(0,1){1}}
\multiput(30,10)(2,0){9}{\line(1,0){1}}
\multiput(30,-2)(0,2){6}{\line(0,1){1}}
\multiput(0,-16)(0,2){9}{\line(0,1){1}}
\put(-1,-20){\small$x'$}
\multiput(-12,0)(2,0){6}{\line(1,0){1}}
\put(-18,-1){\small$n'\alpha'$}
\put(-3,1){\small$\hat r'_1$}
\put(27,12){\small$\hat r'_2$}
\put(45,42){\small$\hat r'_3$}
\put(11,32){\small$\hat r'_4$}
\end{picture}
\end{center}
\begin{quote}
\vspace{-0.5cm}
\caption{\footnotesize An oriented area in phase space defined by $d\mathbf x'=\hat r'_2-\hat r'_1$ and $id(n'\bm\alpha')=\hat r'_3-\hat r'_2$.}
\label{fig:dx dnalpha image e_1ie_2}
\vspace{-0.5cm}
\end{quote}
\end{figure}
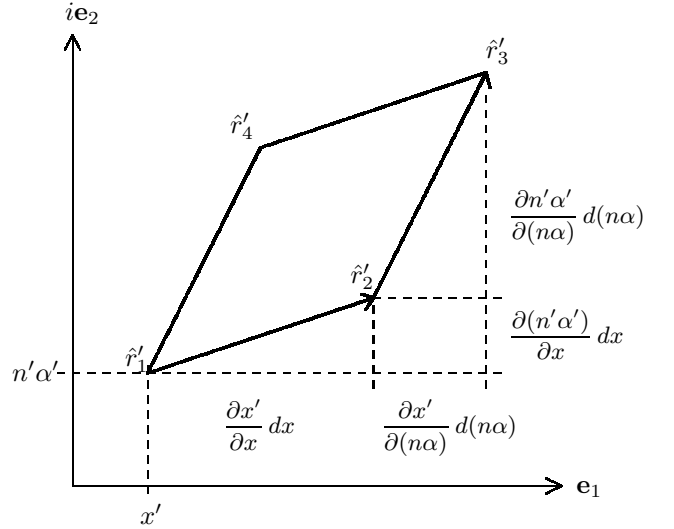

\textbf{b.  Image Phase Rectangle.}  
Since $x'$ and $\alpha'$ are functions of $x$ and $\alpha$, then by the definition of the total differential, we have
\begin{eqnarray}
\label{eq:dx' total differential height angle}
dx'&=&\frac{\partial x'}{\partial x}dx+\frac{\partial x'}{\partial (n\alpha)}d(n\alpha),\\
\label{eq:dn'alpha' total differential height angle}
d(n'\alpha')&=&\frac{\partial(n'\alpha')}{\partial x}dx+\frac{\partial(n'\alpha')}{\partial(n\alpha)}d(n\alpha).
\end{eqnarray}
Using these differential expansions, we can show that the image of the phase rectangle defined by the points $\hat r_1$ to $\hat r_4$ in Eqs.~(\ref{eq:r_1 height angle}) to (\ref{eq:r_4 height angle}) is parallelogram defined by the following points:
\begin{eqnarray}
\label{eq:r'_1 height angle}
\!\!\!\!\hat r'_1&=&x'\mathbf e_1+n'\alpha'\,i\mathbf e_2,\\
\label{eq:r'_2 height angle}
\!\!\!\!\hat r'_2&=&\!\!(x'+\frac{\partial x'}{\partial x}\,dx)\mathbf e_1+(n'\alpha'+\frac{\partial(n'\alpha')}{\partial x}\,dx)i\mathbf e_2,\\
\label{eq:r'_3 height angle}
\!\!\!\!\hat r'_3&=&\!\!(x'+\frac{\partial x'}{\partial x}\,dx+\frac{\partial x'}{\partial(n\alpha)}\,d(n\alpha))\,\mathbf e_1\nonumber\\
& &\!\!\!\!\!\!\!\!+\ (n'\alpha'+\frac{\partial(n'\alpha')}{\partial x}\,dx+\frac{\partial(n'\alpha')}{\partial(n\alpha)}\,d(n\alpha))i\mathbf e_2,\\
\label{eq:r'_4 height angle}
\!\!\!\!\hat r'_4&=&\!\!(x'+\frac{\partial x'}{\partial(n\alpha)}\,d(n\alpha))\,\mathbf e_1\nonumber\\
& &+\  (n'\alpha'+\frac{\partial(n'\alpha')}{\partial(n\alpha)}\,d(n\alpha))i\mathbf e_2.
\end{eqnarray}
Notice that $dx=0$ at points $\hat r_1$ and $\hat r_4$, and $d(n\alpha)=0$ at points $\hat r_1$ and $\hat r_2$; both $dx$ and $d(n\alpha)$ are nonzero at point $\hat r_3$. (Fig. \ref{fig:dx dnalpha image e_1ie_2})

The two consecutive sides of the parallelogram image starting from $\hat r_1$ are
\begin{eqnarray}
\label{eq:r'_12 height angle}
\hat r'_{12}&=&\frac{\partial x'}{\partial x}\,dx\,\mathbf e_1+\frac{\partial(n'\alpha')}{\partial x}\,dx\,i\mathbf e_2,\\
\label{eq:r'_23 height angle}
\hat r'_{23}&=&\frac{\partial x'}{\partial(n\alpha)}\,d(n\alpha)\,\mathbf e_1+\frac{\partial(n'\alpha')}{\partial(n\alpha)}\,d(n\alpha)\,i\mathbf e_2.
\end{eqnarray}
Their product is a scalar-vector cliffor:
\begin{eqnarray}
\label{eq:dr'_12 dr'_23 height angle}
\hat r'_{12}\,\hat r'_{23}\!\!\!\!&=&\!\!\!\!\left(\frac{\partial x'}{\partial x}\frac{\partial x'}{\partial(n\alpha)}-\ \frac{\partial(n'\alpha')}{\partial x}\frac{\partial(n'\alpha')}{\partial(n\alpha)}\right)dx\,d(n\alpha)-\nonumber\\
& &\!\!\!\!\mathbf e_3\left(\frac{\partial x'}{\partial x}\frac{\partial(n'\alpha')}{\partial(n\alpha)}-\frac{\partial x'}{\partial(n\alpha)}\frac{\partial(n'\alpha')}{\partial x}\right) dx\,d(n\alpha).\nonumber\\
\end{eqnarray}
Notice that the magnitude of the vector part is the area of the parallelogram in the image phase space.  The ratio of this area with that of the object in the vector part of Eq.~(\ref{eq:r_12 r_23}) is 
\begin{equation}
\label{eq:Poisson bracket commutator x' n'alpha' wrt x nalpha}
[x',n'\alpha']_{x,n\alpha}=\frac{\partial x'}{\partial x}\frac{\partial(n'\alpha')}{\partial(n\alpha)}-\frac{\partial x'}{\partial(n\alpha)}\frac{\partial(n'\alpha')}{\partial x},
\end{equation}
which is the Poisson commutator bracket.  We shall later show that this commutator is an invariant\cite{Goodman_1995_HandbookofOpticsI_ch1p71} that is equal to unity.

On the other hand, the product of $\hat r_{12}'$ and $\hat r_{23}'^{\,\dagger}$ is
\begin{eqnarray}
\label{eq:dr'_12 dr'_23 dagger height angle}
\hat r'_{12}\,\hat r_{23}'^{\,\dagger}\!\!\!\!&=&\!\!\!\!-\left(\frac{\partial x'}{\partial x}\frac{\partial x'}{\partial(n\alpha)}+ \frac{\partial(n'\alpha')}{\partial x}\frac{\partial(n'\alpha')}{\partial(n\alpha)}\right)dx\,d(n\alpha)-\nonumber\\
& &\!\!\!\!\mathbf e_3\left(\frac{\partial x'}{\partial x}\frac{\partial(n'\alpha')}{\partial(n\alpha)}+\frac{\partial x'}{\partial(n\alpha)}\frac{\partial(n'\alpha')}{\partial x}\right) dx\,d(n\alpha).\nonumber\\
\end{eqnarray}
Since the scalar part does not vanish, then the sides $\hat r_{12}$ and $\hat r_{23}$ of the image parallelogram in phase space are not perpendicular.  For convenience, let us write the scalar part coefficient of $dx\,d(n\alpha)$ as
\begin{equation}
\label{eq:Poisson bracket anticommutator x' n'alpha' wrt x nalpha}
\{x',n'\alpha'\}_{x,n\alpha}=-\left(\frac{\partial x'}{\partial x}\frac{\partial x'}{\partial(n\alpha)}+ \frac{\partial(n'\alpha')}{\partial x}\frac{\partial(n'\alpha')}{\partial(n\alpha)}\right).
\end{equation}
We shall refer to this as a Poisson anticommutator bracket.

\textbf{c.  Symplectic Condition.}  The differentials of Eqs.~(\ref{eq:x' is Ax + Bnalpha}) and (\ref{eq:n'alpha' is -Cx + Dnalpha}) are
\begin{eqnarray}
\label{eq:dx' is A dx + B dnalpha}
dx'&=&A\,dx+B\,d(n\alpha),\\
\label{eq:dn'alpha' is -C dx + D dnalpha}
d(n'\alpha')&=&-C\,dx+D\,n\alpha.
\end{eqnarray}
Comparing these equations with Eqs.~(\ref{eq:dx' total differential height angle}) and (\ref{eq:dn'alpha' total differential height angle}), we arrive at
\begin{eqnarray}
\label{eq:A B as partial derivatives}
A &=&\frac{\partial x'}{\partial x},\qquad\ \ B\ =\ \frac{\partial x'}{\partial(n\alpha)},\\
\label{eq:-C D as partial derivatives}
-C &=&\frac{\partial (n'\alpha')}{\partial x},\quad D\ =\ \frac{\partial(n'\alpha')}{\partial(n\alpha)}.
\end{eqnarray}
Except for the sign of $C$, these equations are the known partial derivative expressions for the system matrix parameters.\cite{Stone_1997_josav14i12pp3415-3429_p3416}

Using the relations in Eqs.~(\ref{eq:A B as partial derivatives}) and (\ref{eq:-C D as partial derivatives}), the Poisson bracket expressions in Eqs.~(\ref{eq:Poisson bracket anticommutator x' n'alpha' wrt x nalpha}) and (\ref{eq:Poisson bracket commutator x' n'alpha' wrt x nalpha}) simplifies to
\begin{eqnarray}
\label{eq:Poisson bracket anticommutator is AB - CD}
\{x',n'\alpha'\}_{x,n\alpha}&=&-(AB-CD),\\
\label{eq:Poisson bracket commutator is AD+BC is 1}
[x',n'\alpha']_{x,n\alpha}&=&AD+BC=1,
\end{eqnarray}
where we used Eq.~(\ref{eq:det M is AD + BC is 1}).  The first equation states that in the phase space, image parallelogram does not preserve the perpendicularity of the sides of the original object parallelogram, while the second equation states that the object and image parallelograms have the same area.  This last condition means that the transformation defined in Eqs.~(\ref{eq:x' is Ax + Bnalpha}) and (\ref{eq:n'alpha' is -Cx + Dnalpha}) is symplectic.

\subsection{Distance-Height Phase Space}

\begin{figure}[ht]
\begin{center}
\setlength{\unitlength}{1 mm}
\begin{picture}(60,60)(-10,-5)
\put(0,0){\line(1,0){45}}
\qbezier(43,-1)(45,0)(45,0)
\qbezier(43,1)(45,0)(45,0)
\put(47,-1){\small$\mathbf e_3$}
\put(0,0){\line(0,1){45}}
\qbezier(-1,43)(0,45)(0,45)
\qbezier(1,43)(0,45)(0,45)
\put(-1,46){\small $\mathbf e_1$}
\multiput(15,0)(0,2){8}{\line(0,1){1}}
\put(12,-5){\small$-S$}
\multiput(0,15)(2,0){8}{\line(1,0){1}}
\put(-5,14){\small$x$}
\thicklines
\multiput(15,15)(0,25){2}{\line(1,0){25}}
\qbezier(39,14)(40,15)(40,15)
\qbezier(39,16)(40,15)(40,15)
\put(27,11){\small$dS$}
\multiput(15,15)(25,0){2}{\line(0,1){25}}
\qbezier(39,39)(40,40)(40,40)
\qbezier(41,39)(40,40)(40,40)
\put(42,25){\small$dx$}
\put(12,11){\small$\mathbf r_1$}
\put(41,11){\small$\mathbf r_2$}
\put(12,41){\small$\mathbf r_4$}
\put(41,41){\small$\mathbf r_3$}
\end{picture}
\end{center}
\begin{quote}
\vspace{-0.5cm}
\caption{\footnotesize An oriented area defined by $dS\,\mathbf e_3$ and $dx\,\mathbf e_1$.}
\label{fig:dS dx e_3e_1}
\vspace{-0.5cm}
\end{quote}
\end{figure}
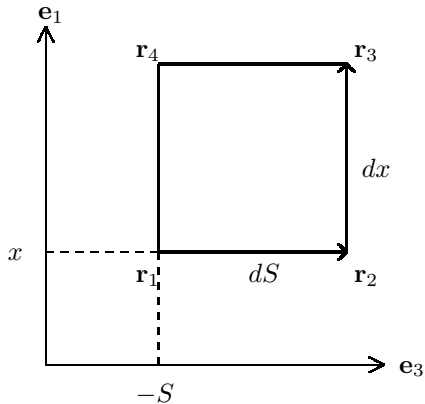

\textbf{a.  Object Rectangle.}  Let us define a rectangular longitudinal object of length $dS$ and height $dx$:
\begin{eqnarray}
\label{eq:r_1 distance height}
\mathbf r_1&=&-S\mathbf e_3+x\mathbf e_1,\\
\label{eq:r_2 distance height}
\mathbf r_2&=&(-S+dS)\mathbf e_3+x\mathbf e_1,\\
\label{eq:r_3 distance height}
\mathbf r_3&=&(-S+dS)\mathbf e_3+(x+dx)\mathbf e_1,\\
\label{eq:r_4 distance height}
\mathbf r_4&=&-S\mathbf e_3+(x+dx)\mathbf e_1,
\end{eqnarray}
where $\mathbf r_1$, $\mathbf r_2$, $\mathbf r_3$, and $\mathbf r_4$ are the four consecutive vertices of the rectangle for a counterclockwise path. (Fig. \ref{fig:dS dx e_3e_1})

The two successive sides of the rectangle are
\begin{eqnarray}
\label{eq:r_12 is dS e_3}
\mathbf r_{12}&=&\mathbf r_2-\mathbf r_1=dS\mathbf e_3,\\
\label{eq:r_23 is dx e_1}
\mathbf r_{23}&=&\mathbf r_3-\mathbf r_2=dx\mathbf e_1.
\end{eqnarray}
Their product is
\begin{equation}
\label{eq:r_12 r_23 is dS dx e_3e_1}
\mathbf r_{12}\mathbf r_{23}=dS\,dx\,\mathbf e_3\mathbf e_1,
\end{equation}
so that
\begin{eqnarray}
\mathbf r_{12}\cdot \mathbf r_{23}&=&0,\\
\label{eq:dr_12 dr_23 bivector is dS dx e_3e_1}
\mathbf r_{12}\wedge \mathbf r_{23}&=&dS\,dx\,\mathbf e_3\mathbf e_1.
\end{eqnarray}
Thus, the sides of the rectangle are perpendicular and the area of the rectangle is $dS\,dx$.

\textbf{a.  Image Parallelogram.}  The differential of $\mathbf r'$ in Eq.~(\ref{eq:r' is S'e_3 + x'e_1}) is 
\begin{equation}
\label{eq:dr' is dS'e_3 + dx'e_1}
d\mathbf r'=dS'\mathbf e_3+dx'\mathbf e_1,
\end{equation}
where
\begin{eqnarray}
\label{eq:dS' total differential}
dS'&=&\frac{\partial S'}{\partial S}dS+\frac{\partial S'}{\partial x}dx,\\
\label{eq:dx' total differential}
dx'&=&\frac{\partial x'}{\partial S}dS+\frac{\partial x'}{\partial x}dx,
\end{eqnarray}
by the definition of a total differential.  

Taking the partial derivatives of $S'$ and $x'$ in Eqs.~(\ref{eq:r' e_3 part transform}) and (\ref{eq:r' e_1 part transform}) with respect to $S$ and $x$, we get
\begin{eqnarray}
\label{eq:partial S' by partial S}
\frac{\partial S'}{\partial S}&=&-\frac{1}{(M_{12}S-M_{22})^2}=-m_x^2,\\
\label{eq:partial S' by partial x}
\frac{\partial S'}{\partial x}&=&0,\\
\label{eq:partial x' by partial S}
\frac{\partial x'}{\partial S}&=&-\frac{M_{12}x}{(M_{12}S-M_{22})^2}=-m_x^2M_{12}x,\\
\label{eq:partial x' by partial x}
\frac{\partial x'}{\partial x}&=&-\frac{1}{(M_{12}S-M_{22})}=m_x,
\end{eqnarray}
where we used the unit determinant property of $\textsf M_{\textrm box}$ in Eq.~(\ref{eq:det M_box is 1}) and the relation for the transverse magnification $m_x$ in Eq.~(\ref{eq:m_x is x' over x}).  Notice that Eq.~(\ref{eq:partial S' by partial S}) is the same as the expression for longitudinal magnification for infinitesimal longitudinal displacement\cite{MouroulisMacDonald_1997_GeometricalOpticsandOpticalDesign_p58}.

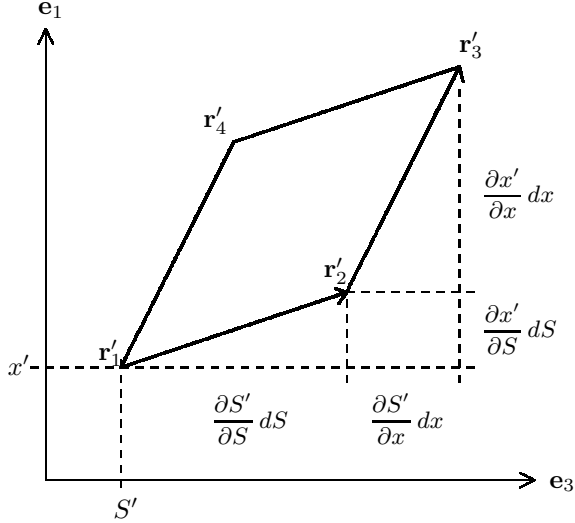
\begin{figure}[hb]
\begin{center}
\setlength{\unitlength}{1 mm}
\begin{picture}(75,85)(-15,-25)
\put(-10,-15){\line(1,0){65}}
\qbezier(53,-14)(55,-15)(55,-15)
\qbezier(53,-16)(55,-15)(55,-15)
\put(57,-16){\small$\mathbf e_3$}
\put(-10,-15){\line(0,1){60}}
\qbezier(-11,43)(-10,45)(-10,45)
\qbezier(-9,43)(-10,45)(-10,45)
\put(-11,47){\small$\mathbf e_1$}

\thicklines
\qbezier(0,0)(30,10)(30,10)
\qbezier(28.5,10.5)(30,10)(30,10)
\qbezier(29,8.5)(30,10)(30,10)

\qbezier(30,10)(45,40)(45,40)
\qbezier(45.5,38)(45,40)(45,40)
\qbezier(0,0)(15,30)(15,30)
\qbezier(15,30)(45,40)(45,40)
\put(12,-8){\footnotesize$\displaystyle\frac{\partial S'}{\partial S}\,dS$}
\put(33,-8){\footnotesize$\displaystyle\frac{\partial S'}{\partial x}\,dx$}
\put(48,4){\footnotesize$\displaystyle\frac{\partial x'}{\partial S}\,dS$}
\put(48,22){\footnotesize$\displaystyle\frac{\partial x'}{\partial x}\,dx$}
\thinlines
\multiput(0,0)(2,0){24}{\line(1,0){1}}
\multiput(45,-2)(0,2){20}{\line(0,1){1}}
\multiput(30,10)(2,0){9}{\line(1,0){1}}
\multiput(30,-2)(0,2){6}{\line(0,1){1}}
\multiput(0,-16)(0,2){9}{\line(0,1){1}}
\put(-1,-20){\small$S'$}
\multiput(-12,0)(2,0){6}{\line(1,0){1}}
\put(-15,-1){\small$x'$}
\put(-3,1){\small$\mathbf r'_1$}
\put(27,12){\small$\mathbf r'_2$}
\put(45,42){\small$\mathbf r'_3$}
\put(11,32){\small$\mathbf r'_4$}
\end{picture}
\end{center}
\begin{quote}
\vspace{-0.5cm}
\caption{\footnotesize The oriented area defined by $d\mathbf x'=\hat r'_2-\hat r'_1$ and $id(n'\bm\alpha')=\hat r'_3-\hat r'_2$.}
\vspace{-0.5cm}
\label{fig:dx dnalpha prime e_3e_1}
\end{quote}
\end{figure}

Now, the image of the points $\mathbf r_1$, $\mathbf r_2$, $\mathbf r_3$, and $\mathbf r_4$ in Eq.~(\ref{eq:r_1 distance height}) to (\ref{eq:r_4 distance height}) are 
\begin{eqnarray}
\label{eq:r'_1 distance height}
\mathbf r'_1&=&S'\mathbf e_3+x'\mathbf e_1,\\
\label{eq:r'_2 distance height}
\mathbf r'_2&=&(S'+\frac{\partial S'}{\partial S}dS)\mathbf e_3+(x'+\frac{\partial x'}{\partial S}dS)\mathbf e_1,\\
\label{eq:r'_3 distance height}
\mathbf r'_3&=&(S'+\frac{\partial S'}{\partial S}dS+\frac{\partial S'}{\partial x}dx)\mathbf e_3\nonumber\\
& &+\ (x'+\frac{\partial x'}{\partial S}dS+\frac{\partial x'}{\partial x}dx)\mathbf e_1,\\
\label{eq:r'_4 distance height}
\mathbf r'_4&=&(S'+\frac{\partial S'}{\partial x}dx)\mathbf e_3+(x'+\frac{\partial x'}{\partial x}dx)\mathbf e_1,
\end{eqnarray}
respectively.  Notice that these points now define a parallelogram.  (Fig.~\ref{fig:dx dnalpha prime e_3e_1})

The two successive sides of the rectangle are
\begin{eqnarray}
\label{eq:r'_12 distance height}
\mathbf r'_{12}&=&\frac{\partial S'}{\partial S}dS\mathbf e_3+\frac{\partial x'}{\partial S}dS\mathbf e_1,\\
\label{eq:r'_23 distance height}
\mathbf r'_{23}&=&\frac{\partial S'}{\partial x}dx\mathbf e_3+\frac{\partial x'}{\partial x}dx\mathbf e_1
\end{eqnarray}
Their product is
\begin{equation}
\label{eq:r'_12 r'_23 distance height}
\mathbf r'_{12}\mathbf r'_{23}=\mathbf r'_{12}\cdot\mathbf r'_{23}+\mathbf r'_{12}\wedge\mathbf r'_{23}
\end{equation}
where
\begin{eqnarray}
\label{eq:r_12 dot r_23 distance height}
\!\!\!\!\!\!\!\!\!\!\!\!\mathbf r'_{12}\!\!\cdot\mathbf r'_{23}\!\!\!\!&=&\!\!\!\!\left(\frac{\partial S'}{\partial S}\frac{\partial S'}{\partial x}+\frac{\partial x'}{\partial S}\frac{\partial x'}{\partial x}\right)dS\,dx,\\
\label{eq:r_12 wedge r_23 distance height}
\!\!\!\!\!\!\!\!\!\!\!\!\mathbf r'_{12}\!\!\wedge\mathbf r'_{23}\!\!\!\!&=&\!\!\!\!\mathbf e_3\mathbf e_1\left(\frac{\partial S'}{\partial S}\frac{\partial x'}{\partial x}-\frac{\partial x'}{\partial S}\frac{\partial S'}{\partial x}\right)dS\,dx
\end{eqnarray}
are the scalar and vector parts.

The coefficient of $dS\,dx$ in Eqs.~(\ref{eq:r_12 dot r_23 distance height}) and (\ref{eq:r_12 wedge r_23 distance height}) may be expressed in terms of the Poisson anticommutator and commutator brackets as
\begin{eqnarray}
\label{eq:Poisson bracket anticommutator S' x'}
\{S',x'\}_{S,x}&=&\frac{\partial S'}{\partial S}\frac{\partial S'}{\partial x}+\frac{\partial x'}{\partial S}\frac{\partial x'}{\partial x},\\
\label{eq:Poisson bracket commutator S' x'}
[S',x']_{S,x}&=&\frac{\partial S'}{\partial S}\frac{\partial x'}{\partial x}-\frac{\partial x'}{\partial S}\frac{\partial S'}{\partial x}.
\end{eqnarray}
Using the magnification equations in Eqs.~(\ref{eq:partial S' by partial S}) to (\ref{eq:partial x' by partial x}), Eqs.~(\ref{eq:Poisson bracket anticommutator S' x'}) and (\ref{eq:Poisson bracket commutator S' x'}) becomes
\begin{eqnarray}
\label{eq:Poisson bracket anticommutator S' x' magnification}
\{S',x'\}_{S,x}&=&-m_x^3M_{12}\,x,\\
\label{eq:Poisson bracket commutator S' x' magnification}
[S',x']_{S,x}&=&-m_x^3.
\end{eqnarray}
The perpendicularity measure in Eq.~(\ref{eq:Poisson bracket anticommutator S' x' magnification}) states that the image of a longitudinal rectangle will also be a rectangle, provided that $M_{12}=0$, which is the characteristic of telescopic systems\cite{KleinFurtak_1986_Optics_p179}.   And the area ratio in Eq.~(\ref{eq:Poisson bracket commutator S' x' magnification}) states that the area of the image is proportional to that of the object by a factor of negative of the cube of the magnification.

\section{Conclusions}

We used geometric algebra to compute the inner (dot) and outer (wedge) product of two vectors.  We used the former to define the Poisson commutator bracket for measuring the perpendicularity of two vectors; the latter for the Poisson anticommutator bracket for measuring areas.  We adopted the complex height-angle ray formalism we developed in previous papers to write down the $2\times 2$ matrix equations for tracing.  And from these equations we derive the partial Moebius transforms that relates the height and distance of the input ray to that of the output ray, as measured from the input and output sides of the optical black box.  

For the case of distance-height rays, we define the object to be a differential rectangle in the $zx-$plane.  We showed that its image  does not preserve the area of the object nor the perpendicularity of its sides.  A special case is that of telescopic systems, where the perpendicularity of the sides is preserved but not the area of the image which is equal to the negative of the cube of the transverse magnification.  The negative sign means that the orientation of the areas is opposite: one area is oriented clockwise; the other counterclockwise.

For the case of the height-angle rays, a similar computation is not possible because we adopted a complex height-angle ray formalism, where height is along the $x-$axis and the angle is along the imaginary $y-$axis.  To define the pependicularity of two complex vectors, we took the scalar part of the product of the complex height-angle ray vector with its spatial inverse: if this scalar part is zero, then the two complex vectors are perpendicular.  To define the area of the phase space parallelogram formed by two complex vectors, we took the magnitude of the vector part of the product of the two complex vectors.  (These procedures are only valid if the vector and imaginary vector components of each vector are perpendicular in the complex sense.)  We showed that the area of differential height-angle rectangular objects in phase space is preserved by a $2\times 2$ matrix transformation; however, the sides of the rectangle are generally not anymore perpendicular.  The ratio of area of the image in phase space to that of the object is equal to the magnitude of the Poisson commutator bracket.

In Differential Geometry, the Poisson commutator bracket is already used and well-known. We also hope that its anticommutator counterpart would also be similary used, within the context of geometric algebra.

\section*{\small{Acknowledgments}}
This research was supported by the Manila Observatory and by the Physics Department of Ateneo de Manila University.

\end{document}